\documentclass[12pt]{JHEP}
\usepackage{epsfig}
\title{On the quantum loop weak interaction corrections at high energies}
\author{D. PALLE\\Zavod za teorijsku fiziku \\ Institut Rugjer
Bo\v skovi\' c\\P. O. Box 180, 10002 Zagreb, CROATIA 
\thanks{This work was supported by the Ministry of Science and
Technology of the Republic of Croatia No. 00980103.}}
\abstract{We perform comparative analyses of quantum loop 
corrections to some observationally important two- and three-point
Green functions within two distinct symmetry-breaking mechanisms.
It appears that the existing high-energy data, neutrino experiments
and present astrophysical and cosmological constraints strongly
disfavour the Higgs mechanism, while the introduction of the
noncontractible space as a symmetry-breaking mechanism can resolve
all known problems and puzzles of fundamental interactions.}

\begin{document}
\section{Introduction and motivation}  
The current and commonly accepted wisdom in particle physics relies
strongly on the formalism of quantum field theory of
unitary gauge symmetries of the Standard Model (SM).
The success of the perturbative calculations and their
agreement with measurements at low and very high energies,
represent the milestone of our confidence into the SM.
However, recent developments in theoretical and
experimental particle physics are far from being considered
satisfactory.

Although there is great discontent with the SM, the SUSY, GUT, etc
extensions of the SM are also based on the Higgs mechanism, thus
preserving all of its bad features: fermion masses are free 
parameters, the introduction of new Higgs scalars to resolve
small neutrino masses, unclear source of fermion mixings, 
production of the large cosmological constant, etc.
Our motivation to change the symmetry-breaking mechanism is
based on the arguments related to the mathematical
inconsistency of the SM \cite{Palle1}, namely the SU(2) global
anomaly and the ultraviolet (UV) singularity.
The mathematically consistent theory (called BY in \cite{Palle1})
violates lepton-number conservation and contains three light
and three heavy Majorana neutrinos, while the finite UV scale
is fixed by the weak interaction scale, explaining simultaneously
broken conformal, gauge and discrete symmetries. 
The dimensionality and noncontractibility of the physical
space should be the only assumption that suffices to unify
strong and electroweak forces 
realized as hidden local symmetries within the
SU(3) conformal scheme.

LSND and SuperKamiokande data refer clearly to the existence 
of massive neutrinos and the neutrino flavour mixing.
Present fits to neutrino data require higher masses and
mixing angles that are close to the estimates in \cite{Palle1}.
Owing to the absence of the Higgs scalars, one is able
to show within the BY theory that heavy neutrinos of
mass ${\cal O}(100 TeV)$ could be candidate particles
as cold dark matter and their lifetimes ${\cal O}(10^{25}s)$ could resolve
the problem of the cosmological diffuse photon
background \cite{Palle2}. 
The first nonvanishing contribution to the CDM particle-nucleon
scattering appears at two loops in the strong coupling and 
DAMA data could be quantitatively understood \cite{DAMA}.

We have also investigated the finite scale effect 
on the running coupling in 
perturbative QCD \cite{Palle3}.
The absence of asymptotic freedom, 
$\lim _{\mu \rightarrow \infty} \alpha_{s}^{\Lambda}(\mu)\ne 0$,
and the enhancement of the strong coupling that starts in
the vicinity of the weak interaction scale, are 
large deviations from the SM.
In this paper we analyse quantum loop weak corrections
and the differences between the SM and the BY theory.
In the next chapter we define renormalization procedures and
in the last chapter we give the results with
remarks and discussions of 
recent high-energy data at colliders.
 
\section{Renormalisation}
In this section we define renormalisation conditions and Green
functions in order to comparatively analyse the SM and the BY
theory.
Since the masses of heavy neutrinos in the BY are at least
a few TeV \cite{Palle2} and the masses of light neutrinos are 
negligible in comparison with charged lepton masses, we
perform the calculations in the BY with massless neutrinos, thus
with no lepton-number violation.

Effectively, we perform the calculations with the Higgs scalar of
the SM and with the UV cut-off and without the Higgs scalar, as in
the BY theory. The spontaneously broken electroweak
theories are renormalisable theories even if there
is no Higgs scalar because in the closed set of 
asymptotic fields that forms the BRST transformations,
the Higgs field is absent \cite{Kugo,Palle1}.

We choose renormalisation conditions for 
the vector gauge boson fields in the
manner to preserve the position and the residue
of the mass singularity of the respective propagator
\cite{Aoki} with particular emphasis on the mixing of 
neutral fields:

\begin{eqnarray}
\Delta (q^{2})^{-1} \equiv q^{2} - M_{V}^{2} +
 \Sigma^{R} (q^{2}), 
\end{eqnarray}
\begin{eqnarray*}
\Sigma^{R}_{i}(M_{i}^{2})=0,\ \frac{d\Sigma^{R}_{i}}{dq^{2}}
(q^{2}=M_{i}^{2})=0,
\end{eqnarray*}
\begin{eqnarray*}
\Sigma^{R}_{i}(q^{2}) = \Sigma^{U}_{i}(q^{2})+              
\Sigma^{C}_{i}(q^{2}),\ i=Z,W, 
\end{eqnarray*}
\begin{eqnarray*}
\Sigma^{C}_{i}(q^{2}) = \delta M^{2}_{i} +
Z_{i} (M^{2}_{i}-q^{2}), 
\end{eqnarray*}
\begin{eqnarray*}
\Sigma^{U}_{i}(M^{2}_{i}) + \delta M^{2}_{i}=0, 
\end{eqnarray*}
\begin{eqnarray}
\frac{d\Sigma^{U}_{i}}{dq^{2}}(q^{2}=M^{2}_{i}) - Z_{i}=0, 
\end{eqnarray}
\begin{eqnarray*}
\Sigma^{R}_{ZA}(0)=0,\ \Sigma^{R}_{ZA}(M_{Z}^{2})=0,
\end{eqnarray*}
\begin{eqnarray*}          
\Sigma^{R}_{ZA}(q^{2}) = \Sigma^{U}_{ZA}(q^{2}) +
\Sigma^{C}_{ZA}(q^{2}), 
\end{eqnarray*}
\begin{eqnarray*}
\Sigma^{ZA}_{C}(q^{2}) = Z^{1/2}_{ZA} (M^{2}_{Z} - q^{2})
- Z^{1/2}_{AZ} q^{2},
\end{eqnarray*}
\begin{eqnarray*}
\Sigma^{U}_{ZA} (M^{2}_{Z}) - M^{2}_{Z} Z^{1/2}_{AZ} = 0 ,
\end{eqnarray*}
\begin{eqnarray}
\Sigma^{U}_{ZA} (0) + M^{2}_{Z} Z^{1/2}_{ZA} = 0 , 
\end{eqnarray}
\begin{eqnarray*}
\Sigma^{R}_{A}(0)=0, \frac{d\Sigma^{R}_{A}}{dq^{2}}(q^{2}=0)=0,
\end{eqnarray*}
\begin{eqnarray*}    
\Sigma^{R}_{A} (q^{2}) = \Sigma^{U}_{A} (q^{2}) +
\Sigma^{C}_{A} (q^{2}), 
\end{eqnarray*}
\begin{eqnarray*}
\Sigma^{C}_{A} (q^{2}) = -2 q^{2} Z^{1/2}_{AA}, 
\end{eqnarray*}
\begin{eqnarray}
\frac{d\Sigma^{U}_{A}}{dq^{2}}(q^{2}=0) - 2 Z^{1/2}_{AA} = 0 . 
\end{eqnarray}

Similarly, one has to impose renormalisation conditions on the fermion
propagators with mixing  \cite{Gambino} and the most natural
choice for the mixing operators is
$\Sigma^{R}_{ij}(m_{i}^{2})=0$ and $\Sigma^{R}_{ij}(m_{j}^{2})=0$,
i,j=flavour indices.
In the calculation of the boson propagators we neglect the fermion
mixing effect, because it is numerically unimportant.

We have to mention that the above renorm conditions 
differ markedly from those in Ref.\cite{Boehm}.
The conditions of B\"{o}hm et al do not fulfil
the requirements for the propagator to match the free-field mass
singularity structure.
It is not clear how their renorm conditions can remove UV
singularity in the structure $C_{UV}^{\infty}(k^{2}-M_{V}^{2})$
of the propagator.
 As a consequence, one cannot use 
their renormalised propagators directly in the  
evaluation of the observables, such as the effective weak
mixing angle. 
The result of the unsuitable renorm conditions in \cite{Boehm} 
is the appearance of a completely spurious term for
the quantum correction to the effective weak-mixing
angle \cite{Hollik}

\begin{eqnarray}
M^{2}_{W} s^{2}_{W} &=& \frac{\pi\alpha_{e}}{\sqrt{2}G_{\mu}}
\frac{1}{1-\bigtriangleup r}, \nonumber \\
\bigtriangleup r &=& \frac{\Sigma^{R}_{W}(0)}{M^{2}_{W}}
+ \frac{\alpha_{e}}{4\pi s^{2}_{W}}
(6 + \frac{7-4 s^{2}_{W}}{2 s^{2}_{W}}\ln c^{2}_{W}),
\nonumber \\
(\bigtriangleup r)_{top} &\approx & -\frac{\alpha_{e}}{4\pi}
\frac{3 c^{2}_{W}}{4 s^{4}_{W}}\frac{m^{2}_{t}}
{M^{2}_{W}}, \\
s_{W} &\equiv &\sin \theta_{W},\ c_{W} \equiv \cos \theta_{W}.
\nonumber
\end{eqnarray}

Further numerical comparison of two renorm schemes is given
in the next section.

It is very well known that the on-shell renorm conditions 
to the electron-photon vertex remove any further
divergences of the electroweak vertices \cite{Aoki,Boehm}.
However, we choose two conditions for the
vertices under our study \cite{Hollik}:

\begin{eqnarray}
F^{Zf}_{V}(0)_{weak} = F^{Zf}_{A}(0)_{weak} = 0.
\end{eqnarray}

For our purpose,
it will suffice to compare quantum
corrections for the SM and the BY theory. 

Although the renorm conditions for the SM and the BY theory
should be the same, the differences between the renorm
procedures are (1) presence (absence)
of the Higgs scalar, (2) absence (presence) of the UV cut-off
in the scalar Green functions.
The Green functions with the UV cut-off should preserve the following 
characteristics: 
(1) the real parts should be defined after Wick's rotation and
integration in the spacelike region up to the covariant cut-off
and analytically continued on the Riemann sheets above
the threshold in the timelike region, matching the standard
Green functions in the limit $\Lambda=\infty$,
(2) because of the broken scale invariance for $\Lambda < \infty$,
one has to symmetrise the Green function over the masses and
external momenta to preserve the exchange symmetry properties 
of the standard regularised functions (see Appendix B).
 
\section{Results and discussion}
Following the procedures described in the preceding section, one can
evaluate renormalised gauge boson propagators in the SM and 
the BY theory (see Appendix A for the unrenormalised functions).

At first place, in Fig. 1 we display Z and W renorm propagators
of the SM in order to make a comparison with the 
renorm scheme of B\" {o}hm et al (see Figs. 10 and 11 of
Ref. \cite{Boehm}).

One can see that the difference between the two schemes
 is substantial and the 
contribution of the correct W gauge boson renorm propagator
to the effective weak-mixing angle is dominated by the
gauge boson loops without any enhancement due to
the heavy top quark mass $\bigtriangleup r = -0.0413$
(parameters of this paper,see below).

With the set of parameters

\begin{eqnarray*}
M_{H}&=&200 GeV,\ M_{W}=80.44 GeV,\ M_{Z}=91.19 GeV,\ m_{s}=0.16 GeV, \\
m_{c}&=&2 GeV,\ m_{b}=4.5 GeV,\ m_{t}=175 GeV,\ \Lambda=326 GeV,
\end{eqnarray*}

in Fig. 2 we draw the Z-boson renormalised propagators of the SM and the BY 
theory to make a direct comparison (dependence on the Higgs scalar
mass does not influence essentially the result).
 
\EPSFIGURE[b]{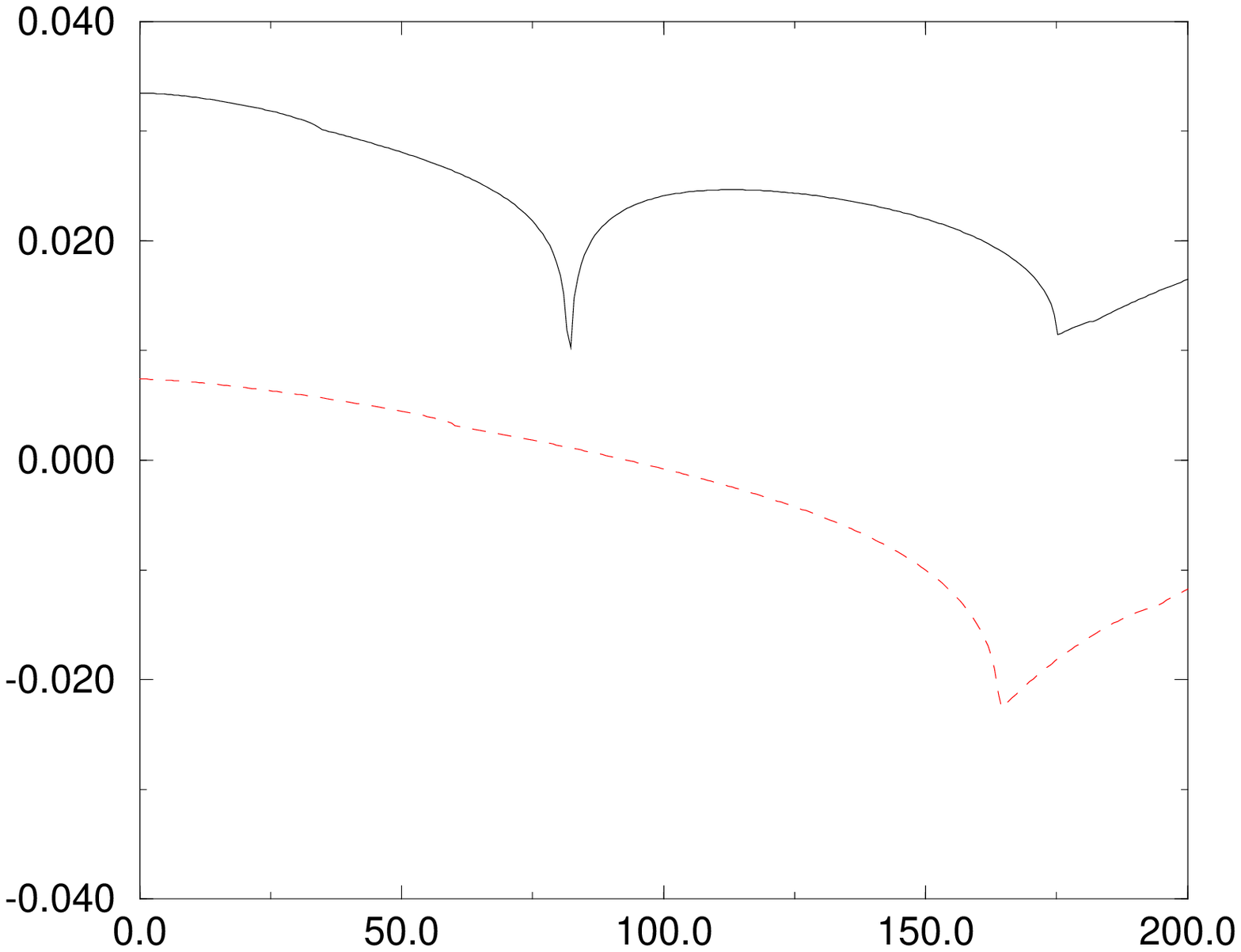,width=140mm,height=120mm}{Solid [dashed] line denotes 
$\Sigma^{R}_{W}(p^{2})/(p^{2}-M_{W}^{2})$
 [$\Sigma^{R}_{Z}(p^{2})/(p^{2}-M_{Z}^{2})$] vs. p(GeV);
parameters as in Ref. \cite{Boehm}. }

\EPSFIGURE[t]{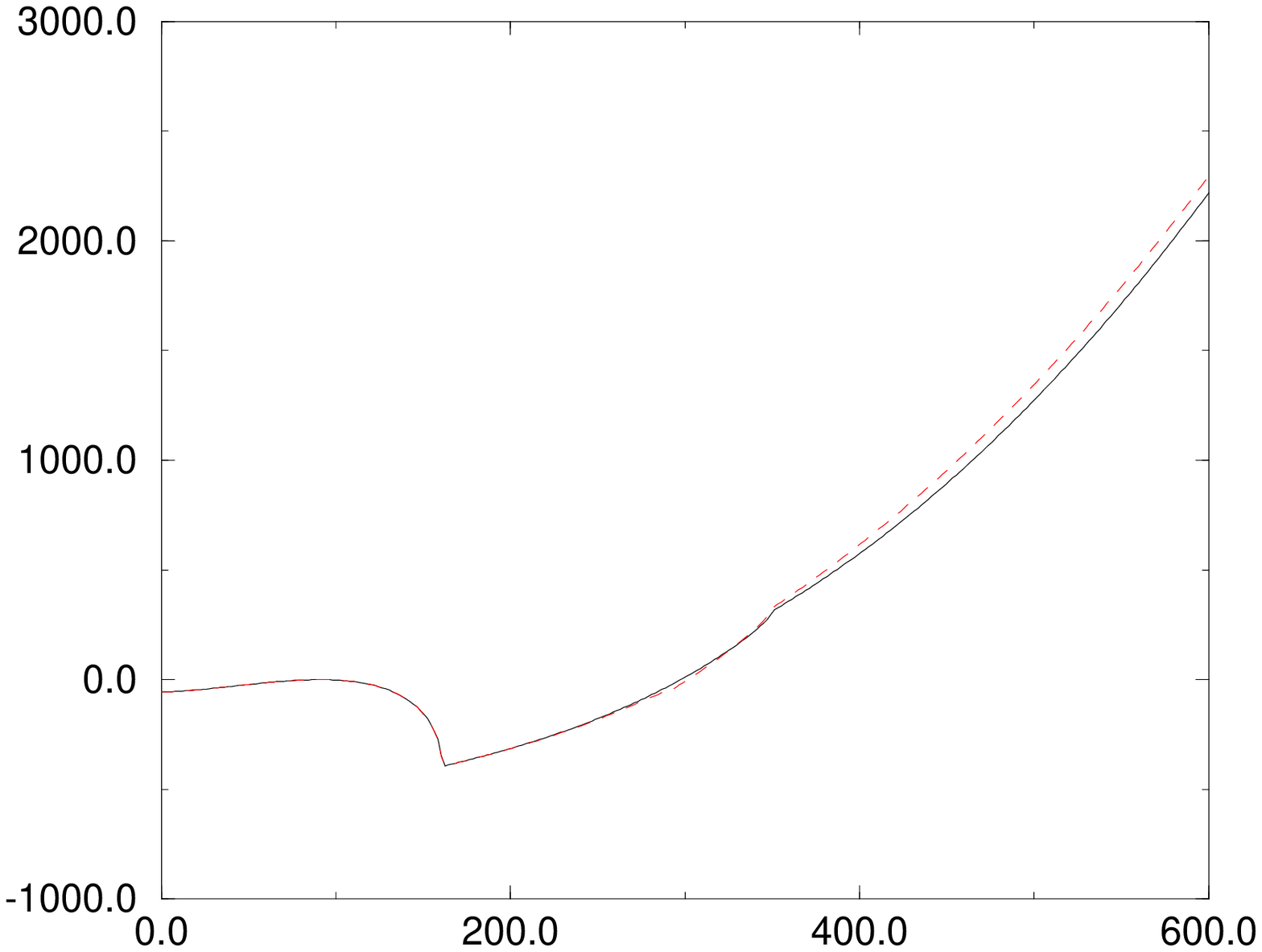,width=140mm,height=140mm}{Solid [dashed] line denotes
$\Sigma^{R}_{Z}(p^{2})_{\Lambda}$ 
[$\Sigma^{R}_{Z}(p^{2})_{\infty}$](GeV$^{2}$) vs. p (GeV). }

One can notice that a
substantial 
difference appears and grows starting from the scale $p\simeq 
400 GeV$. 

Recently it has been reported that there is a deviation of
the weak charge SM prediction from the
measurements of parity-nonconservation in Cs \cite{Bennett}.
The BY theory evidently cannot improve the situation; however,
it seems that an additional atomic-structure calculation can
explain this discrepancy  \cite{Derevianko}.

In a similar fashion we compare quantum loop weak
corrections to the vector and axial-vector couplings
of heavy quarks. With the unrenormalised vertices of Append. A
and Green functions of Append. B, the renorm procedure
leads us to the following results:

\begin{eqnarray}
v_{f}&=&v_{f}^{0}+F^{Zf}_{V}(weak)+..., \\
a_{f}&=&a_{f}^{0}+F^{Zf}_{A}(weak)+..., \\
v_{f}^{0}&=&(I^{3}_{f}-2 s^{2}_{W} Q_{f})/2s_{W}c_{W},\ 
a_{f}^{0}=I^{3}_{f}/2s_{W}c_{W}. \nonumber
\end{eqnarray}

\TABULAR[h]{|l||l|l|l|l|l|l|l|l|} 
{\hline
$\sqrt{s}$(GeV) & 100 & 150 & 200 & 250 & 400 & 550 & 700 & 850 \\
\hline \hline
$10^{3}\cdot (F^{Zc}_{V}(\Lambda)-F^{Zc}_{V}(\infty))$&0.44&0.49&0.60&0.77
&5.60&7.75&12.29&19.10 \\
\hline 
$10^{3}\cdot (F^{Zc}_{A}(\Lambda)-F^{Zc}_{A}(\infty))$&0.43&0.47&0.58&0.74&5.39&
7.28&11.83&18.16 \\
\hline }
{Weak corrections to the Z-c quark vertex.}

\TABULAR[h]{|l||l|l|l|l|l|l|l|l|} 
{\hline
$\sqrt{s}$(GeV) & 100 & 150 & 200 & 250 & 400 & 550 & 700 & 850 \\
\hline \hline
$10^{3}\cdot (F^{Zb}_{V}(\Lambda)-F^{Zb}_{V}(\infty))$&-1.95&-0.93&-0.56&-0.38
&-0.28&-1.43&-6.56&-7.71 \\
\hline 
$10^{3}\cdot (F^{Zb}_{A}(\Lambda)-F^{Zb}_{A}(\infty))$&-1.19&-0.61&-0.38&-0.26&-0.19&
-1.98&-7.89&-8.65 \\
\hline }
{Weak corrections to the Z-b quark vertex.}

\TABULAR[h]{|l||l|l|l|l|} 
{\hline
$\sqrt{s}$(GeV) & 400 & 550 & 700 & 850 \\
\hline \hline
$10^{3}\cdot (F^{Zt}_{V}(\Lambda)-F^{Zt}_{V}(\infty))$&-2.31&1.08&12.15
&16.03 \\
\hline 
$10^{3}\cdot (F^{Zt}_{A}(\Lambda)-F^{Zt}_{A}(\infty))$&-6.30&-1.88&11.90
&16.15 \\
\hline }
{Weak corrections to the Z-t quark vertex.}

Loop corrections to the weak couplings are two orders of
magnitude smaller than the tree level values, and
significant differences between the SM and the BY 
grow after the scale $\sqrt{s}\simeq 400 GeV$.
One should also remember that any effective electroweak vertex  
with quarks also
contains the QCD correction factor (higher-order
corrections could be found in \cite{Chetyrkin})

\begin{eqnarray*}
|V_{EW+QCD}(q^{2})|^{2} = |V_{EW}(q^{2})|^{2} [1 + \frac{\alpha_{s}(q^{2})}
{\pi} + {\cal O}(\alpha^{2}_{s}(q^{2}))] .
\end{eqnarray*}

The presented numerical results, together with our 
previous work, allow us to make a concluding
discussion and observations:
(1) the difference between the SM and BY  weak
corrections to the weak couplings of heavy quarks 
becomes effective at the rather large scale 
$\sqrt{s} \simeq 400 GeV$;
(2) the observed nonresonant enhancement at HERA 
could be attributed to the QCD enhacement factor
$(1+\frac{\alpha_{s}^{\Lambda}(p^{2})}{\pi})/(1+\frac{\alpha_{s}^{\infty}
(p^{2})}{\pi})$
to the squared electroweak couplings of quarks \cite{HERA}
starting at $p\simeq 200 GeV$;
(3) a similar effect one expects at lepton colliders only for
the jet production channel; this nonresonant enhancement is  probably 
observed at LEP 2 \cite{Tully};
(4) a clear signal at hadron high-energy colliders should
come from the enhancement of the amplitude 
owing to the factor $\alpha_{s}^{\Lambda}(\mu)/
\alpha_{s}^{\infty}(\mu)$ \cite{Tevatron1};
one could easily check that parton distributions are not
very much affected by the stronger $\alpha_{s}$: there
is a very small enhancement for small x and a very
small suppression for large x \cite{Kumano}, thus
it is very difficult to find it in experimental data;
(5) Run 2 of TeVatron, together with a new HERA run,
are capable to resolve the existence and nature
of this QCD effect;
(6) it is not excluded that deviations of the electroweak
couplings could be measured at TeVatron \cite{Tevatron2};
(7) it has been shown that the Einstein-Cartan nonsingular cosmology
can solve the problem of the cosmic mass density,
cosmological constant problem  \cite{Palle4} and the primordial 
mass density fluctuation \cite{Palle5} without the
introduction of the scalar (inflaton) field;
(8) to conclude, one can say that the theory of vacuum and
the Higgs mechanism is
like a modern theory of ether, and it is natural to
expect that Nature should choose only a mathematically
consistent theory to describe the physical laws.

\section{Appendix A}
In this appendix we summarise unrenormalised gauge
boson self energies and unrenormalised vector and
axial-vector Z-boson-heavy quark vertices ('t Hooft-Feynman gauge):

\begin{eqnarray*}
\Sigma^{U}_{A}(s)&=&\frac{\alpha_{e}}{4\pi}[\frac{4}{3}\sum_{f}
Q_{f}^{2}(s+2 m_{f}^{2})(B_{0}(s;m_{f},m_{f}) 
-B_{0}(0;m_{f},m_{f})) \\ &-&(3s+4M_{W}^{2})(B_{0}(s;M_{W},M_{W})-
B_{0}(0;M_{W},M_{W}))] , 
\end{eqnarray*}
\begin{eqnarray*}
\Sigma^{U}_{W}(s)&=&\frac{\alpha_{e}}{4\pi}\frac{1}{s_{W}^{2}}
\sum_{f,f'}[2B_{22}-s B_{1}](s;m_{f},m_{f'}) \\
&+&\frac{\alpha_{e}}{4\pi}[-(1+10\frac{c_{W}^{2}}{s_{W}^{2}})
B_{22}(s;M_{Z},M_{W})-\frac{1}{s_{W}^{2}}B_{22}(s;M_{H},M_{W}) \\
&-&9B_{22}(s;0,M_{W})+\frac{c_{W}^{2}}{s_{W}^{2}}
B_{22}(s;M_{W},M_{Z})+B_{22}(s;M_{W},0) \\
&-&\frac{M_{W}^{2}}{s_{W}^{2}}B_{0}(s;M_{H},M_{W})
-(M_{W}^{2}+5 s) B_{0}(s;0,M_{W})\\ &-&(\frac{s_{W}^{2}}{c_{W}^{2}} 
M_{W}^{2}-2 \frac{c_{W}^{2}}{s_{W}^{2}}M_{Z}^{2} 
+5 \frac{c_{W}^{2}}{s_{W}^{2}} s)B_{0}(s;M_{Z},M_{W}) \\ 
&-&2 s B_{1}(s;0,M_{W})-2\frac{c_{W}^{2}}{s_{W}^{2}} s
B_{1}(s;M_{Z},M_{W})]+terms\ with\ A(m_{i}) ,
\end{eqnarray*}
\begin{eqnarray*}
\Sigma^{U}_{Z}(s)&=&\frac{\alpha_{e}}{4\pi} 4\sum_{f}(v_{f}^{2}
+a_{f}^{2})(2 B_{22}(s;m_{f},m_{f})-s B_{1}(s;m_{f},m_{f})) \\
&-&\frac{\alpha_{e}}{4\pi}\frac{1}{c_{W}^{2}s_{W}^{2}}
[B_{22}(s;M_{H},M_{Z})+M_{Z}^{2} B_{0}(s;M_{H},M_{Z})] \\
&-&\frac{\alpha_{e}}{4\pi}[\frac{(s_{W}^{2}-c_{W}^{2})^{2}}
{4c_{W}^{2}s_{W}^{2}} 4B_{22}+2\frac{s_{W}^{2}}{s_{W}^{2}}
M_{W}^{2}B_{0} \\
&+&\frac{c_{W}^{2}}{s_{W}^{2}}(2 M_{W}^{2} B_{0}+
8 B_{22}+5 s B_{0}+2 s B_{1})](s;M_{W},M_{W}) \\
&+&\ terms\ with\ A(m_{i}) ,
\end{eqnarray*}   
\begin{eqnarray*}
\Sigma^{U}_{AZ}(s)&=&\frac{\alpha_{e}}{4\pi}[-\sum_{f}
4v_{f}Q_{f}(2B_{22}-sB_{1})(s;m_{f},m_{f}) \\
&+&[\frac{c_{W}}{s_{W}}(12 B_{22}+(5 s-4 M_{W}^{2})B_{0}
+2 s B_{1})+\frac{1}{c_{W}s_{W}}(-2 B_{22}+
2 M_{W}^{2}B_{0})](s;M_{W},M_{W})] \\
&+&\ terms\ with\ A(m_{i}).
\end{eqnarray*}

Expressions for weak vertices are as in Ref. \cite{Beenakker} Table 2,
except: with defined $g^{\pm}_{f}\equiv v_{f}\pm a_{f}$, one
reads for diagram (k) $g^{\pm}_{f}$ instead of $g^{\mp}_{f}$ and
for diagram (o) the third column changes a sign. All 
relevant definitions could be found in Ref. \cite{Beenakker}.

\section{Appendix B}
Here we define  two- and three-point scalar functions for
the SM (\cite{scalar}) and the BY theory:

\begin{eqnarray*}
B_{0}(p^{2};m_{1},m_{2})=\frac{1}{\imath\pi^{2}}\int d^{4}q
\frac{1}{(q^{2}-m_{1}^{2}+\imath\varepsilon)
((q+p)^{2}-m_{2}^{2}+\imath\varepsilon)},  \\
\end{eqnarray*}
\begin{eqnarray*}
p_{\mu}B_{1}(p^{2};m_{1},m_{2})=\frac{1}{\imath\pi^{2}}
\int d^{4}q\frac{q_{\mu}}{(q^{2}-m_{1}^{2}+
\imath\varepsilon)((q+p)^{2}-m_{2}^{2}+\imath\varepsilon)},  \\
\end{eqnarray*}
\begin{eqnarray*}
g_{\mu\nu}B_{22}+p_{\mu}p_{\nu}B_{21}=
\frac{1}{\imath\pi^{2}}
\int d^{4}q\frac{q_{\mu}q_{\nu}}{(q^{2}-m_{1}^{2}+
\imath\varepsilon)((q+p)^{2}-m_{2}^{2}+\imath\varepsilon)}, \\
\end{eqnarray*}
\begin{eqnarray*}
C_{0}(p_{1}^{2},p_{2}^{2},p_{3}^{2};m_{1},m_{2},m_{3})=
\frac{1}{\imath\pi^{2}}\int 
\frac{d^{4}q}{(q^{2}-m_{1}^{2}+\imath\varepsilon)((q+p_{1})^{2}-m_{2}^{2}+
\imath\varepsilon)((q-p_{3})^{2}-m_{3}^{2}+\imath\varepsilon)}, \\
\end{eqnarray*}

The real parts of the two- and three-point scalar Green
functions in the noncontractible space are given
as in Ref. \cite{Palle3}(effects of symmetrisation now included):

\begin{eqnarray*}
\Re B_{0}^{\Lambda}(p^{2};m_{1},m_{2})=
\frac{1}{2}[\Re \tilde{B}_{0}^{\Lambda}(p^{2};m_{1},m_{2})+
\Re \tilde{B}_{0}^{\Lambda}(p^{2};m_{2},m_{1})],
\end{eqnarray*}
\begin{eqnarray*}
\Re \tilde{B}_{0}^{\Lambda}(p^{2};m_{1},m_{2})=
(\int_{0}^{\Lambda^{2}}d y K(p^{2},y)+\theta (p^{2}-m_{2}^{2}) 
 \int_{-(\sqrt{p^{2}}-m_{2})^{2}}
^{0}d y \Delta K(p^{2},y) )\frac{1}{y+m_{1}^{2}}, \\
K(p^{2},y)=\frac{2 y}{-p^{2}+y+m_{2}^{2}+
\sqrt{(-p^{2}+y+m_{2}^{2})^{2}+4 p^{2} y}},  \hspace*{40 mm}\\
\Delta K(p^{2},y)=\frac{\sqrt{(-p^{2}+y+m_{2}^{2})^{2}+4 p^{2} y}}{p^{2}}.
 \hspace*{60 mm}
\end{eqnarray*}

The integration in the second term is performed from the branch 
point of the square root $\sqrt{(-p^{2}+y+m_{2}^{2})^{2}+4 p^{2} y}\equiv 
\imath Z$ and the additional kernel is derived as the difference:
$ \Delta K(p^{2},y)=K(p^{2},y)-K^{*}(p^{2},y)=\frac{2 y}{-p^{2}+y+
m_{2}^{2}+\imath Z}-\frac{2 y}{-p^{2}+y+m_{2}^{2}-\imath Z}$.

The integration over singularities is supposed to be the principal-value 
integration.

In the case of the two-point Green function $B_{0}^{\Lambda}$,
 we need the explicit form of the additional term for the integration
in the timelike region because the integration 
in the spacelike region is divergent in the limes $\Lambda 
 \rightarrow \infty$. However, the three-point scalar Green
functions are UV-convergent and we do not need to know the explicit
form of the additional terms because they do not depend on the 
UV cut-off and we can use the analytical continuation of the standard
 Green functions written in terms of the dilogarithms\cite{Palle3,scalar}:

\begin{eqnarray*}
\Re C_{0}^{\Lambda}(p_{1}^{2},p_{2}^{2},p_{3}^{2};
m_{1}^{2},m_{2}^{2},m_{3}^{2}) &=& \frac{1}{3}
[\Re \tilde{C}_{0}^{\Lambda}(p_{1}^{2},p_{2}^{2},p_{3}^{2};
m_{1}^{2},m_{2}^{2},m_{3}^{2}) \\
&+& \Re \tilde{C}_{0}^{\Lambda}(p_{2}^{2},p_{3}^{2},p_{1}^{2};
m_{2}^{2},m_{3}^{2},m_{1}^{2})+
\Re \tilde{C}_{0}^{\Lambda}(p_{3}^{2},p_{1}^{2},p_{2}^{2};
m_{3}^{2},m_{1}^{2},m_{2}^{2}],
\end{eqnarray*}
\begin{eqnarray*}
\Re C_{0}^{\Lambda}(p_{i},m_{j})=\int_{0}^{\Lambda^{2}}dq^{2} \Phi 
(q^{2},p_{i},m_{j})
+\int_{TD}dq^{2} \Xi (q^{2},p_{i},m_{j}), \hspace*{10 mm}\\
\Re C_{0}^{\Lambda}(p_{i},m_{j})=Re C_{0}^{\infty}(p_{i},m_{j})-
\int^{\infty}_{\Lambda^{2}}d q^{2} \Phi (q^{2},p_{i},m_{j}), 
 \hspace*{20 mm} \\
\Phi \equiv function\ derived\ by\ the\ angular\ integration\ 
after\ Wick's\ rotation,\\
C_{0}^{\infty}\equiv standard\ 't\ Hooft-Veltman\ scalar\ function, 
\hspace*{20 mm} \\
TD\equiv timelike\ domain\ of\ integration. \hspace*{40 mm}
\end{eqnarray*}
\begin{eqnarray*}
\Re C^{\Lambda}=\Re C^{\infty} - \bigtriangleup \Gamma,
\end{eqnarray*}
\begin{eqnarray*}
\bigtriangleup \Gamma = \Gamma ^{\infty}-\Gamma ^{\Lambda}.
\end{eqnarray*}

This equation is valid for arbitrary external momenta. The same formula
 is applicable to the higher n-point one loop scalar Green functions
(procedure could be generalised to multiloop Green functions).

We need the following functions:
 
\begin{eqnarray*}
\tilde{\Gamma}^{\Lambda}_{123}&=&-\frac{2}{\pi}\frac{1}{\sqrt{s(s-4m_{q}^{2})}}
\int^{\Lambda}_{0}dq\frac{q}{q^{2}+m_{1}^2} \\
&\times &\int^{+1}_{-1}\frac{dx}{x}\frac{1}{2}\sum^{3}_{i=2}
(\arctan\frac{A_{i}+B}{\sqrt{s}qx}-\arctan\frac{A_{i}-B}{\sqrt{s}qx}),\\
A_{i}&=&q^{2}+m_{q}^{2}+m_{i}^{2},\ B=q\sqrt{1-x^{2}}\sqrt{s-4m_{q}^{2}}, 
\end{eqnarray*}
\begin{eqnarray*}
p_{1}^{2}=m_{f}^{2},\ p_{2}^{2}=s,\ p_{3}^{2}=m_{f}^{2},
\end{eqnarray*}
\begin{eqnarray*}   
\tilde{\Gamma}^{\Lambda}_{231}\ and\ \tilde{\Gamma}^{\Lambda}_{312}\ 
are\ evaluated\ in\ the\ same\ manner.
\end{eqnarray*}

The integration for high $s=k^{2}$ could be performed 
with sufficient accuracy
only with Monte Carlo Riemannian integration.

Further examples of Green functions are:

\begin{eqnarray*}
C^{-}_{1}(s=0)=\frac{1}{2}(m_{2}^{2}-m_{1}^{2})\frac{\partial C_{0}}
{\partial s}(s=0),
\end{eqnarray*}
\begin{eqnarray*}
C_{0}(s=0,m_{3}=m_{q})=\frac{1}{m_{2}^{2}-m_{1}^{2}}
\int^{1}_{0}dx\ln \frac{m_{q}^{2}x^{2}-m_{1}^{2}x+m_{1}^{2}}
{m_{q}^{2}x^{2}-m_{2}^{2}x+m_{2}^{2}}.
\end{eqnarray*}

One can easily evaluate $\frac{dB_{0}^{\Lambda;\infty}}{ds} < \infty$ from its integral
representation, for instance:

\begin{eqnarray*}
\frac{d\tilde{B}_{0}^{\Lambda}}{ds}(s=0;m_{1},m_{2})&=&
\frac{m_{1}^{2}m_{2}^{2}}{(m_{1}^{2}-m_{2}^{2})^{3}}
\ln \frac{m_{2}^{2}(\Lambda^{2}+m_{1}^{2})}
{m_{1}^{2}(\Lambda^{2}+m_{2}^{2})}\\
&+&\frac{1}{(m_{1}^{2}-m_{2}^{2})^{2}(\Lambda^{2}+m_{2}^{2})^{2}}
(\frac{m_{1}^{2}+m_{2}^{2}}{2}\Lambda^{4}+m_{2}^{4}\Lambda^{2}), 
\end{eqnarray*}
\begin{eqnarray*}
B_{1}(0)=\frac{1}{2}[-B_{0}(0)+(m_{2}^{2}-m_{1}^{2})B_{0}'(0)],
\end{eqnarray*}
\begin{eqnarray*}
B_{1}'(0)=-\frac{1}{2}B_{0}'(0)+\frac{1}{4}(m_{2}^{2}-
m_{1}^{2})B_{0}''(0) .
\end{eqnarray*}

\end{document}